\documentclass[conference]{IEEEtran}
\IEEEoverridecommandlockouts

\usepackage{cite}
\usepackage{amsmath,amssymb,amsfonts}
\usepackage{algorithmic}
\usepackage{graphicx}
\usepackage{textcomp}
\usepackage{xcolor}
\def\BibTeX{{\rm B\kern-.05em{\sc i\kern-.025em b}\kern-.08em
    T\kern-.1667em\lower.7ex\hbox{E}\kern-.125emX}}
\begin{document}

\title{Socio-technical and Ethical Dimensions of Architecture Practices in FLOSS
\thanks{\textcopyright 2026 IEEE. Reproduced with permission.\\Published in: 2026 IEEE International Conference on Software Maintenance and Evolution (ICSME)}
}

\author{
	\IEEEauthorblockN{Sven Thielen}
	\IEEEauthorblockA{
		\textit{Faculty of Mathematics and Natural Sciences}\\
		\textit{Heinrich Heine University Düsseldorf}\\
		Düsseldorf, Germany\\
		0009-0004-3487-9137
	}
}

\maketitle

\begin{abstract}
This project investigates how software architecture practices in Free/Libre Open Source Software (FLOSS) are shaped by socio-technical and ethical factors, and how education can support more explicit, inclusive, and reflective architectural work. Motivated by FLOSS's role in digital sovereignty, it is  observed that architectural decisions are often undocumented and scattered across issues, pull requests, and mailing lists. While prior research has studied architectural artifacts, erosion, and communication, the interplay between architectural work, governance arrangements, and ethical commitments in FLOSS remains underexplored.
	
The research follows a three-phase design: (1) multi-method case studies of 3--4 domain-pairs of architecturally non-trivial FLOSS projects, (2) framework and intervention design with practitioners and educators, and (3) pilot evaluations in projects and courses. It will produce (i) cross-case empirical evidence on FLOSS architecture practices, (ii) a conceptual framework linking architecture practices to socio-technical conditions and ethical dimensions, and (iii) lightweight practices and teaching formats that render architectural work more explicit and inclusive.
\end{abstract}


\section{Introduction}
\label{sec:INT}

Free/Libre Open Source Software (FLOSS) constitutes critical infrastructure and is frequently linked to transparency, autonomy, and sustainability. Maintaining these systems therefore depends on the conditions under which architectural decisions are made, communicated, and sustained. Many FLOSS projects rely on emergent, informal architecture; rationale and constraints are only partly captured in artifacts and are scattered across issues, pull requests (PRs), and mailing lists \cite{migliorini2024archviews,bi2021archinfo}. Prior work highlights architectural erosion \cite{le2018archdecay,thielen2025from} and a concentration of decision authority in a few maintainers \cite{pinheiro2024group}, which creates bottlenecks for review, onboarding, and change.

This project investigates how FLOSS architecture practices are shaped by socio-technical and ethical factors, and how can FLOSS-based education support more explicit and reflective work? We conduct multi-method case studies, followed by the design and formative evaluation of lightweight practices and teaching formats. The expected contributions are (i) cross-case empirical evidence, (ii) a conceptual framework, and (iii) actionable, low-overhead practices and teaching formats.

\section{Background and Research Gap}
\label{sec:BRG}

Research on software architecture in FLOSS provides relevant insights, but it is fragmented across distinct strands that rarely integrate architectural practice with governance and ethical concerns in an integrated way. The following paragraphs summarize the most relevant strands for this project.

\textbf{(S1) Architectural artifacts, recovery, and erosion.} Empirical studies report that FLOSS projects often rely on partially documented or implicit architectural knowledge, and that architectural erosion can threaten maintainability and long-term evolution \cite{migliorini2024archviews,le2018archdecay,thielen2025from}. This strand motivates studying what architectural knowledge is captured (e.g., design documents, architecture decision records (ADRs), module boundaries) and how it changes over time.

\textbf{(S2) Communication of architectural knowledge.} Architectural rationale in FLOSS is frequently distributed across heterogeneous channels (e.g., issues, pull requests, mailing lists), making it difficult to locate, reuse, and transfer \cite{bi2021archinfo}. Complementary analyses of practitioner discourse indicate that architecture discussions may emphasize later-stage concerns (e.g., deployment and operations), leaving earlier decision-making and rationale underrepresented \cite{su2026emerging}. This motivates examining where architectural decisions are negotiated and how rationale becomes (or fails to become) durable project knowledge.

\textbf{(S3) Governance, roles, and decision points.} FLOSS governance research has characterized mechanisms of formalization, decentralization, and control in commons-based peer production \cite{rozas2021loosenctrl}. Work on maintainership and review pathways (e.g., in the Linux kernel) suggests that responsibility distribution can be operationalized through concrete role arrangements that shape who can approve and coordinate changes \cite{pinheiro2024group}. However, governance studies rarely connect these mechanisms explicitly to architectural work (e.g., how architectural boundaries are negotiated, who sets constraints, and how architectural stewardship is maintained).

\textbf{(S4) Ethical concerns, ecosystem health, and participation.} Ethical notions such as stewardship, responsibility, and inclusivity are frequently invoked in FLOSS. Architectural practices have been argued to influence ecosystem health \cite{amorim2023ecosystemhealth}. However, these concepts are seldom studied at the level of day-to-day architectural work in projects, and they are rarely operationalized into observable indicators (e.g., participation pathways in architecture-relevant discussions, traceability of decision rationale, or newcomer access to architectural knowledge).

\textbf{(S5) FLOSS in education and architecture training.} FLOSS projects are increasingly used in software engineering education to provide authentic, large-scale systems and real collaboration contexts \cite{silva2019floss,lessa2020approach}. At the same time, software architecture education emphasizes the need to teach trade-offs and decision-making beyond purely technical competencies \cite{pantoja2024training}. However, it remains unclear to what extent FLOSS-based teaching explicitly addresses architectural practices as socio-technical work, including questions of responsibility distribution, participation, and rationale capture.

\textbf{Research gap.} While strands (S1)--(S5) provide valuable perspectives, it remains underexplored how architectural work in FLOSS emerges at the intersection of artifacts, communication routines, and governance arrangements, and how this intersection relates to ethical concerns such as stewardship, responsibility, and inclusivity. In addition, there is limited evidence on lightweight, actionable practices that improve the durability and accessibility of architectural knowledge without imposing substantial overhead on maintainers, and on how such practices can be translated into educational formats.

\textbf{Project aim.} To address this gap, a multi-method case studies of FLOSS projects will be conducted, a conceptual framework linking architectural practice to socio-technical and ethical dimensions will be developed, and lightweight interventions for practice and education will be designed and formatively evaluated.

\section{Research Questions and Objectives}
\label{sec:REQ}

The research design is intentionally iterative: early findings will inform refinement and refocusing of subsequent questions.

\noindent\textbf{Overarching research question:} \emph{How are FLOSS architecture practices shaped by socio-technical and ethical factors, and how can FLOSS-based education make this work more explicit, inclusive and reflective?}

\subsection{Sub-questions}
\label{subsec:SUQ}

\begin{itemize}
	\item[RQ1] How do selected projects document, communicate and coordinate architecture, and how do these practices evolve?
	\item[RQ2] How do governance, communication structures and ethical considerations shape participation and rationale preservation?
	\item[RQ3] How do FLOSS-based courses teach architectural decision-making and related socio-technical responsibilities?
	\item[RQ4] Which lightweight practices and educational formats can improve explicitness and accessibility without raising maintainer workload?
\end{itemize}

These questions are expected to evolve as empirical insights accumulate. The educational strand is therefore treated as a core component of this project rather than an optional add-on. The design of teaching formats directly stems from the empirical findings and is evaluated with the same rigor as the practice-oriented interventions. Table~\ref{tab:S2D} maps the research strands to the objectives.

\subsection{Objectives}
\label{subsec:OBJ}

\begin{itemize}
	\item[O1:] Systematically characterize architecture artifacts, communication channels, and decision-making mechanisms in a set of FLOSS projects.
	\item[O2:] Analyze how socio-technical and ethical factors shape architectural practices and knowledge distribution in these projects.
	\item[O3:] Examine how FLOSS-based teaching currently presents software architecture and whether it addresses socio-technical and ethical aspects.
	\item[O4:] Develop a conceptual framework that links architecture practices, socio-technical conditions, and ethical considerations in FLOSS.
	\item[O5:] Design and preliminarily evaluate lightweight architectural practices and educational patterns that support more explicit and reflective architecture work in FLOSS.
\end{itemize}

\begin{table}[htbp]
	\caption{From state-of-the-art strands to dissertation objectives}
	\label{tab:S2D}
	\centering
	\scriptsize
	\begin{tabular*}{\columnwidth}{@{\extracolsep{\fill}} p{0.42\linewidth} p{0.52\linewidth}@{}}
		\hline
		\textbf{Strand} & \textbf{Corresponding Objective(s)} \\
		\hline
		(S1) Architectural artifacts, recovery, erosion &
		O1: systematic characterization of artifacts and their evolution \\
		(S2) Communication of architectural knowledge &
		O1: capture of communication routines; \newline O2: analyze how they shape knowledge flow \\
		(S3) Governance, roles, decision points &
		O2: link governance structures to decision authority; \newline O4: feed into design of interventions \\
		(S4) Ethical concerns, ecosystem health, participation &
		O2: identify ethical dimensions (stewardship, inclusivity); \newline O4: translate into lightweight practices \\
		(S5) FLOSS in education &
		O3: analyse current teaching practice; \newline O5: create and evaluate educational formats \\
		\hline
	\end{tabular*}
\end{table}

\section{Methodology}
\label{sec:MET}

The study follows three iterative phases:

\begin{enumerate}
	\item Information gathering (repo mining, communication analysis, interviews);
	\item Solution generation (framework and lightweight practice design);
	\item Evaluation (pilots in projects and courses).
\end{enumerate}

Repository mining, communication analysis, and semi-structured interviews will be combined to capture both observable architectural artifacts and the tacit knowledge that shapes decision-making, thereby avoiding the treatment of social and technical aspects in isolation \cite{hoda2022stgt}. Educational materials will be analyzed to examine how FLOSS-based teaching currently represents architecture practices and whether the socio-technical and ethical dimensions are sufficiently addresses. In Table~\ref{tab:R2D}, each research question is linked to the primary data sources and the expected methodological outputs (e.g., recovered architectural views, coded decision rationale, and cross-case themes).

\begin{table}[htbp]
	\caption{Mapping research questions to data sources and methods.}
	\label{tab:R2D}
	\centering
	\scriptsize
	\begin{center}
		\begin{tabular*}{\columnwidth}{@{\extracolsep{\fill}} p{0.04\linewidth} p{0.38\linewidth} p{0.48\linewidth}@{}}
		\hline
		\textbf{RQ} & \textbf{Data} & \textbf{Method / Output} \\
		\hline
		RQ1 & Repos, ADRs/docs, dependency structure, issues/PRs & Mining + architectural recovery; description of artifacts and coordination patterns \\
		RQ2 & Governance docs, communication threads, interviews & Qualitative coding + case comparison; factors shaping responsibility/participation \\
		RQ3 & Syllabi, assignments, teaching cases & Document analysis; characterization of how architecture and ethics are addressed \\
		RQ4 & Co-design feedback, pilot observations & Design + formative evaluation; lightweight practices and educational patterns \\
		\hline
		\end{tabular*}
	\end{center}
\end{table}

\begin{figure*}[htbp]
	\centering
	\includegraphics[width=\textwidth]{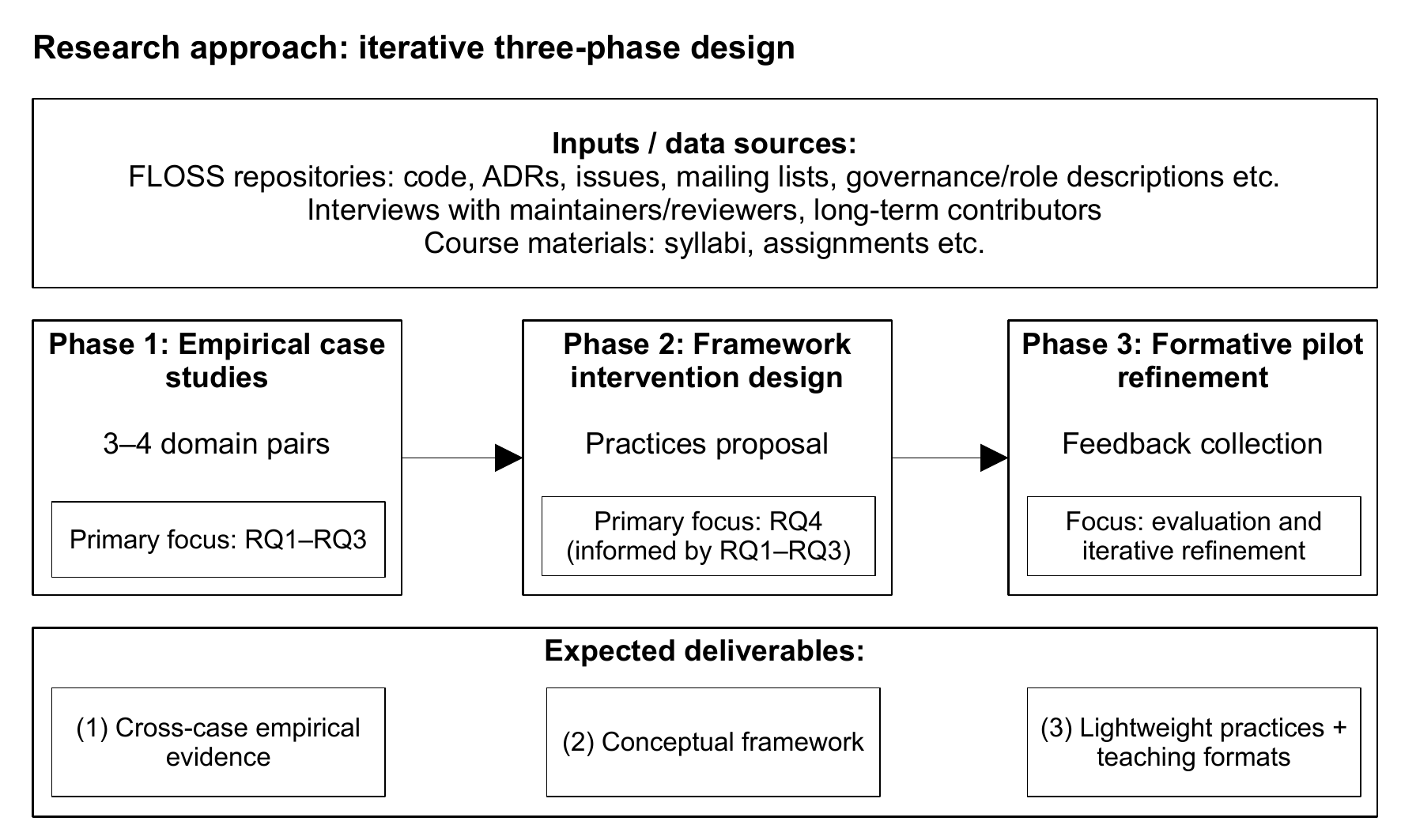}
	\caption{Overview of the iterative three-phase research design.}
	\label{fig:RAV}
\end{figure*}

Each case will produce (i) a description of the project's architectural knowledge base (artifacts and their traceability), (ii) a map of decision points and participation pathways (who proposes, discusses, and approves architecture-relevant changes), and (iii) themes describing socio-technical and ethical factors shaping architectural work. Fig.~\ref{fig:RAV} provides a visual overview of the three-phase research design.

\subsection{Phase 1: Information Gathering and Interpretation}
\label{subsec:PH1}

(1) \textbf{Literature review:} a systematic review of FLOSS architecture practice, architectural erosion/smells, architecture knowledge communication, and FLOSS-based education will be conducted.

(2) \textbf{Case selection:} 3--4 domain-pairs are examined, each pair sharing a comparable architectural core while differing in governance. Selection depends on accessibility of repositories and communication channels. Candidate projects include:

\begin{itemize}
	\item Office suites: LibreOffice vs. OpenOffice.org,
	\item Cryptographic toolchains: GnuPG vs. Sequoia-PGP,
	\item Package managers: Pacman vs. Zypper,
	\item Messaging ecosystems: Matrix ecosystem vs. SimpleX,
	\item Theorem provers: Rocq vs. Z3.
\end{itemize}	

Inclusion/exclusion criteria will be reviewed and refined after the initial literature scan and a pilot inspection of artifacts and communication traces; additional criteria (e.g., traceability of decisions, feasibility of recruiting interviewees, suitability for educational activities) will be documented transparently.

From each repository, architecture-related artifacts (module structures, design documents) will be extracted and dependency graphs will be computed. Architectural recovery (static‑analysis and clustering) will yield views of coupling and modularity. Where possible quantitative proxies will be derived for the five analytic dimensions introduced in the conceptual model:

\begin{enumerate}
	\item Governance: Presence of formal governance documents will indicate whether an explicit architectural authority exists. Role-based permission matrices (GitHub or GitLab team settings) will reveal the concentration of push/merge rights on architecture‑relevant files.
	\item Participation: The number of unique contributors that appear in architecture-tagged threads will be counted and the ratio of newcomers (first-time contributors) to veteran participants in those threads will be computed.
	\item Decision-power distribution: Ownership concentration will be measured on file-ownership for architectural modules. In parallel, it will be counted how many reviewers have ever approved an ADR or a design-critical pull request.
	\item Ethical stewardship: Design documents and ADRs will be scanned for explicit mentions of ethical concerns (privacy, sustainability, inclusivity). The existence of ``participation pathways’’ (e.g., mentorship tags or ``good-first-issue’’ labels) will be recorded.
	\item Inclusivity: Diversity of role-types (developers, translators, UX designers, documentation contributors, etc.) will be extracted from mailing-list participant metadata. As a supporting indicator, sentiment or politeness scores will be computed for architecture-related discussions.
\end{enumerate}

All proxies will be obtained automatically where feasible (e.g., ownership metrics from the version-control history) and will be validated through the interview phase. The combination of quantitative indicators and qualitative interview data provides a transparent triangulation for reasoning about latent socio-technical and ethical constructs.

Issues, pull requests and mailing-list archives will be mined for architecture‑relevant threads. Threads are identified through keyword filtering, label‑based metadata, and explicit links from ADRs or design documents to the discussion artifacts. A stratified manual sample will validate the retrieval, and the inclusion/exclusion criterion for what constitutes an ``architectural decision'' or ``architectural constraint'' will be refined iteratively per project. 10--20 semi-structured interviews with maintainers, reviewers and long-term contributors will be conducted to explore how they conceptualize and enact architecture work and what challenges they perceive. A small corpus of course descriptions, syllabi, and teaching materials from FLOSS-oriented software engineering courses will be collected.

The results will provide (i) a descriptive account of current practices (RQ1), (ii) initial insights into the socio-technical and ethical factors (RQ2), and (iii) a baseline of how architecture is taught in FLOSS-based courses (RQ3). These findings will guide the refinement of research questions, sampling strategies, and data collection instruments for the subsequent phases.

\subsection{ Phase 2: Solution Generation Based on Findings}
\label{subsec:PH2}

Based on the patterns identified in Phase 1, a framework that relates observed architectural practices to socio-technical factors (e.g., governance, communication structures, roles) and to ethical considerations (e.g., inclusivity, responsibility, stewardship) will be constructed. Lightweight architectural practices for FLOSS projects (e.g., adapted templates for documenting decisions, guidelines for involving newcomers in architecture discussions, or practices for making architectural constraints visible) will be proposed \cite{amorim2023ecosystemhealth}. Prototype educational formats or resources (e.g., course assignments, reflective exercises, or teaching cases) that foreground architecture as a socio-technical and ethical practice will be developed. A subset of FLOSS contributors and educators will be engaged through co-design sessions or feedback interviews to refine these practices and educational formats. Phase 2 focuses on RQ4, using empirical findings and stakeholder input to generate context-sensitive and realistic solutions.

\subsection{Phase 3: Evaluation of Proposed Solutions}
\label{subsec:PH3}

In the final phase, the proposed practices and educational formats will be evaluated and refined. Selected architectural practices (e.g., documentation templates or meeting structures) will be piloted in one or two collaborating FLOSS projects. Educational prototypes will be integrated into one or more teaching settings (e.g., a FLOSS-oriented software engineering course). Qualitative feedback (e.g., interviews, questionnaires, reflexive reports) from contributors and students on feasibility, perceived usefulness, and effects on understanding and participation will be collected. Lightweight quantitative cues will be tracked before and after the pilots, such as (i) adoption and completion rates of the proposed decision/rationale templates, (ii) traceability links between ADRs/design discussions and code changes (e.g., ADR~$\leftrightarrow$~PR references), and (iii) participation in architecture-relevant discussions (e.g., number of unique contributors and newcomer participation in architecture-tagged threads). Given the formative nature of the pilots, these cues will be treated as indicators to guide refinement rather than as evidence for causal effects \cite{pantoja2024training}. Evaluation results will be used to further refine the conceptual framework and practical guidelines. Phase 3 strengthens the practical relevance and validity of the research outcomes and supports the development of actionable recommendations for FLOSS communities and educators.

\subsection{Validity, Trustworthiness, and Ethics}
\label{subsec:VTE}

To enhance the robustness of findings, the study will apply the following measures:

\begin{itemize}
	\item \textbf{Triangulation:} Multiple data sources (repositories, artifacts, communication logs, interviews, and educational materials) will be combined and converging and diverging evidence will be compared across sources \cite{gunatilake2024empathy}.
	\item \textbf{Audit trail and transparency:} A coding diary will be maintained, sampling and inclusion/exclusion decisions (e.g., for architecture-relevant threads) will be documented, and the evolving codebook and analysis scripts will be versioned. Where feasible, scripts and de-identified/aggregated datasets will be shared \cite{hoda2022stgt}.
	\item \textbf{Peer debriefing:} Coding decisions, emerging themes, and alternative interpretations will be discussed periodically with the supervisory team or research peers (without implying co-authorship) to challenge assumptions and reduce researcher blind spots.
	\item \textbf{Member checking (interviews):} Short summaries of interpreted themes will be shared with interview participants to confirm plausibility and to reveal misunderstandings.
	\item \textbf{Negative case analysis:} Counterexamples will be actively sought within and across cases (e.g., projects where architectural knowledge is explicit but participation remains concentrated, or vice versa) and explanations will be refined accordingly.
	\item \textbf{Reflexivity:} The researcher's position and potential biases will be reflected upon when interpreting socio-technical and ethical dimensions.
\end{itemize}

\textbf{Risks and mitigation.} The most critical risks stem from (i) limited access to maintainers for interviews, (ii) incomplete traceability of architectural rationale, (iii) the possibility that the selected case set does not capture the full spectrum of governance models, and (iv) the rapidly evolving role of generative AI/LLMs in software development. Mitigation strategies are summarised in Table~\ref{tab:KRM}.

\begin{table}[htbp]
	\caption{Key risks and planned mitigation actions}
	\label{tab:KRM}
	\centering
	\scriptsize
	\begin{tabular*}{\columnwidth}{@{\extracolsep{\fill}} p{0.34\linewidth} p{0.58\linewidth}@{}}
		\hline
		\textbf{Risk} & \textbf{Mitigation}\\
		\hline
		Interview non-response & Early outreach via project maintainers; offer flexible interview formats (e-mail, video, async).\\
		Missing architectural rationale & Combine keyword-based trace mining with manual validation; use interview data to fill gaps.\\
		Insufficient domain coverage & Use the paired-case strategy (two projects per domain) to guarantee at least two instances per governance type.\\
		Obscured authorship & Detect auto-generated commits (e.g., presence of ``generated-by-ChatGPT'' in commit messages) and treat them as a separate ``AI-assisted'' provenance class during ownership analysis.\\
		Scope creep & Adopt an iterative ``stop-criterion'': if after the first two pairs the emerging patterns already saturate (no new themes in coding), subsequent pairs will be used mainly for validation rather than exploration.\\
		\hline
	\end{tabular*}
\end{table}

\textbf{AI/LLM note.} Recent advances in generative AI are already being used in contributions. While this does not become a primary focus of this work, it will be recorded whether an architectural decision or an ADR was produced (or edited) with the assistance of an LLM (e.g., as identified in commit messages). This lightweight capture allows to assess any systematic impact of AI on decision authority and knowledge distribution without expanding the scope of this project.

\textbf{Additional risk: interpreting communication traces.} Text-based communication can be perceived very differently by different developers, increasing the risk of misclassification when relying on trace data alone \cite{herrmann2025perception}. Therefore, sensitive interpretations (e.g., conflict) will be triangulated via interviews and reported cautiously.

\textbf{Ethical considerations.} For interviews, informed consent will be obtained and participants can withdraw at any time. Recordings/transcripts will be stored securely and reported in aggregated form; direct quotes will be paraphrased or de-identified where necessary to reduce re-identification risks. For trace data, the study will avoid attributing sensitive interpretations to individuals and will anonymize project- and person-level details where appropriate.

\section{Expected Results}
\label{sec:EXP}

The expected outcome is a grounded account of how architectural practices are enacted in FLOSS communities, together with a conceptual framework and pragmatic guidance that can support more explicit and reflective architectural work. By connecting empirical software engineering with educational design, the project aims to produce insights that are relevant both to researchers and to FLOSS practitioners. It complements practitioner-oriented perspectives with a more holistic lens on architectural design, quality, and evolution \cite{su2026emerging}. Expected deliverables:

\begin{itemize}
	\item \textbf{Empirical evidence:} cross-case insights into how architectural practices are carried out and communicated in FLOSS projects (artifacts, communication patterns, recurring challenges).
	\item \textbf{Explanatory analysis:} how roles, governance, and community norms shape knowledge distribution, participation, and architectural decision-making.
	\item \textbf{Conceptual framework:} links between architecture practices, socio-technical conditions, and ethical considerations in FLOSS.
	\item \textbf{Interventions:} lightweight practices and educational patterns supporting explicit, inclusive, and reflective architecture work in FLOSS and in FLOSS-based teaching.
	\item \textbf{Research outputs:} publications and (where possible) open research data and scripts (e.g., via Zenodo), respecting licensing and ethical constraints.
\end{itemize}

\section{Research Plan and Milestones}
\label{sec:RPM}

Table~\ref{tab:PTL} outlines the planned milestones. The schedule may be refined based on access to projects and early empirical findings.

\begin{table}[htbp]
	\caption{Planned timeline and milestones.}
	\label{tab:PTL}
	\centering
	\scriptsize
	\begin{tabular*}{\columnwidth}{@{\extracolsep{\fill}} p{0.22\linewidth} p{0.70\linewidth}@{}}
		\hline
		\textbf{Time} & \textbf{Milestones} \\
		\hline
		2026 Q3--Q4 & Finalize case selection; finalize data collection protocol; pilot extraction of artifacts and discussion threads. \\
		2027 Q1--Q2 & Within-case analyses: repository mining + communication analysis; begin interviews; draft case reports. \\
		2027 Q3 & Cross-case synthesis; draft conceptual framework (socio-technical and ethical dimensions). \\
		2027 Q4--2028 Q1 & Co-design and refinement of lightweight practices and educational formats; prepare pilot deployments. \\
		2028 Q2 & Formative evaluations in 1--2 projects and 1--2 courses; iterate on framework and interventions. \\
		2028 Q3--Q4 & Consolidate results; dissertation writing and submission. \\
		\hline
	\end{tabular*}
\end{table}

\section{Feedback Sought}
\label{sec:FBS}

This project seeks feedback from the Doctoral Symposium on the following points:

\begin{itemize}
	\item \textbf{Operationalization:} Which proxies and analysis strategies are defensible to characterize architectural responsibility and decision power distribution from FLOSS traces?
	\item \textbf{Case strategy:} Is the planned case pairing and diversity appropriate in terms of scope and for cross-case synthesis, or is a single case per domain sufficient and should cases be narrowed?
	\item \textbf{Interventions:} Which lightweight interventions are realistic in FLOSS practice and education without increasing maintainer workload (and how should feasibility be assessed)?
\end{itemize}

\section*{Acknowledgment}
During manuscript preparation, the author used ChatGPT (OpenAI) for language-editing assistance (e.g., phrasing and grammar suggestions) and to explore alternative structures for clarity. The author takes full responsibility for the content and verified all technical claims and references.

\end{document}